\documentclass[prl,superscriptaddress,preprint,endfloats]{revtex4}

\newcommand {\EC}{EuCd$_2${}}
\newcommand {\CC}{CeCu$_2${}}
\newcommand {\AB}{AlB$_2${}}
\newcommand {\AO}{Al$_2$O$_3${}}
\newcommand {\GPS}{Gd$_2$PdSi$_3${}}

\newcommand {\TN}{$T_\mathrm{N}${}}

\newcommand {\sigmaxx}{$\sigma _{xx}${}}

\newcommand {\rhoxx}{$\rho _{xx}${}}
\newcommand {\rhoyx}{$\rho _{yx}${}}
\newcommand {\rhoyxOHE}{$\rho _{yx}^{\mathrm{O}}${}}
\newcommand {\rhoyxAHE}{$\rho _{yx}^{\mathrm{A}}${}}
\newcommand {\rhoyxTHE}{$\rho _{yx}^{\mathrm{T}}${}}
\newcommand {\Deltarhoyx}{$\varDelta\rho _{yx}${}}
\newcommand {\DeltarhoyxAHE}{$\varDelta\rho _{yx}^{\mathrm{A}}${}}
\newcommand {\DeltarhoyxTHE}{$\varDelta\rho _{yx}^{\mathrm{T}}${}}

\newcommand{\bm}[1]{{\mbox{\boldmath $#1$}}}
\newcommand {\Bin}{$B_{\mathrm{in}}${}}
\newcommand {\Bout}{$B_{\mathrm{out}}${}}
\newcommand {\BFC}{$B_{\mathrm{FC}}${}}
\usepackage{graphicx}
\usepackage{bm} 
\usepackage{pifont}
\usepackage{amsmath, amsthm, amssymb}
\usepackage{color}
\usepackage{dcolumn} 
\begin{document}
\title{Topological Hall effect enhanced at magnetic transition fields\\ in a frustrated magnet \EC}
\author{S. Nishihaya}
\affiliation{Department of Physics, Tokyo Institute of Technology, Tokyo 152-8551, Japan}
\author{Y. Watanabe}
\affiliation{Department of Physics, Tokyo Institute of Technology, Tokyo 152-8551, Japan}
\author{M. Kriener}
\affiliation{RIKEN Center for Emergent Matter Science (CEMS), Wako 351-0198, Japan}
\author{A. Nakamura}
\affiliation{Department of Physics, Tokyo Institute of Technology, Tokyo 152-8551, Japan}
\author{M. Uchida}
\email[Author to whom correspondence should be addressed: ]{m.uchida@phys.titech.ac.jp}
\affiliation{Department of Physics, Tokyo Institute of Technology, Tokyo 152-8551, Japan}

\begin{abstract}
Emergent magnetic fields exerted by topological spin textures of magnets lead to an additional Hall response of itinerant carriers called the topological Hall effect (THE). While THE as a bulk effect has been widely studied, THE driven by magnetic domain boundaries (DBs) has been elusive. Here, we report rich Hall responses characterized by multiple peak structures and a hysteresis loop in films of EuCd$_2$, where Eu layers form a geometrically frustrated lattice of Heisenberg spins. We uncover a THE component sharply enhanced at magnetic transition fields, indicating a giant contribution from non-trivial spin textures possibly formed at the DBs.
\end{abstract}
\maketitle

Hall effects which are neither proportional to the magnetic field nor to the magnetization have been one of the most vital research topics surrounding quantum transport phenomena in condensed-matter physics. Such non-monotonic Hall responses have been associated mostly with a non-monotonic modulation of the Berry phase rooted in either momentum space or real space. The momentum-space case originates from drastic changes of the band structure accompanied by the formation or shift of singular band features such as Weyl points during the magnetization process, and can be understood in the same framework as the intrinsic anomalous Hall effect (AHE) \cite{AHE,AHE_scaling}. On the other hand, real-space-based Berry phase is induced when itinerant carriers couple with non-coplanar magnetic orderings \cite{THE,Nd2Mo2O7}. As exemplified by skyrmion phases realized in chiral and geometrically-frustrated magnets \cite{MnSi,MnGe,FeGe,GdPdSi,EuAl4}, a non-coplanar arrangement of spins is characterized by the scalar spin chirality defined by the solid angle spanned by three spins ($\bm{S}_{1}\cdot \bm{S}_{2}\times \bm{S}_{3}$). The scalar spin chirality acts as an emergent magnetic field on charged carriers, leading to additional contributions to the Hall effect usually termed as topological Hall effect (THE). In addition to the above-mentioned Berry-phase origins, recent studies have also revealed that the extrinsic skew scattering can lead to an additional giant Hall response when combined with local scalar spin chirality of fluctuating spins, overcoming the upper limit of the Hall conductivity set by the intrinsic band structure \cite{Chiralityscattering,MnGe_scattering,EuAs}. Thus, the observation of the non-monotonic AHE or THE provides experimental evidence of unique electronic and magnetic structures as well as novel scattering mechanisms.    

While the non-monotonic AHE and THE as bulk effects within a single magnetic domain have been widely reported in various systems, the observation of Hall contributions from a domain boundary (DB) has been rare. DBs in magnets generally serve as an additional scattering source or conduction path depending on the details of the magnetic orderings \cite{DWscattering,NIO,EIO}, and are expected to give finite contributions to the Hall responses when coupled with a non-zero Berry phase. In ferromagnets, for example, conventional DBs between the ferromagnetic domains exhibit topologically trivial spin textures leading to no THE, whereas chiral domain walls (skyrmion bubbles) induced by interfacial Dzyaloshinskii-Moriya (DM) interaction in films and heterointerfaces, or topological defects such as vertical Bloch lines are predicted to contribute to finite THE \cite{Bubble,DWTHE}. When the system possesses Weyl points near the Fermi level, enhanced skew-scattering of Weyl fermions at the DB is proposed to lead to a giant AHE even comparable to the bulk contribution \cite{DWWeyl}. The recent observation of a non-monotonic AHE accompanied by a unique hysteresis loop in the ferromagnetic Weyl semimetal CeAlSi and related compounds have been attributed to the scattering at DBs \cite{CeAlSi_1,CeAlSi_2,CeAlGe2}. However, DB contributions in magnets with more complex spin orderings such as non-colinear or non-coplanar configurations have been elusive, except for one known example of pyrochlore-type antiferromagnets where the all-in-all-out ordering forms conductive DBs responsible for unconventional AHE \cite{DWAHE}. The DBs in systems with complex magnetic orderings can lead to unique spin configurations formed locally, and thus serve as a fertile playground for exploring novel Hall responses.

In this Letter, we demonstrate a giant THE triggered by magnetic transitions in films of a frustrated magnet \EC . \EC\ takes the \CC -type hexagonal structure, which can be viewed as a lower-symmetry derivative of the \AB -type structure with an alternate stacking of triangular and honeycomb lattices \cite{AlB2_strc}. In the \CC -type structure of \EC , the Eu triangular lattice is anisotropically distorted in the $ac$-plane and the Cd honeycomb lattice is buckled along the $b$-axis as shown in Fig. 1. Several \AB- and \CC -type compounds with Heisenberg-type spins located at the triangular sites have been known to host non-colinear/non-coplanar orderings such as the skyrmion phase in Gd$_2$PdSi$_3$ \cite{GdPdSi}, the spiral spin texture in EuZnGe \cite{EuZnGe}, or the non-colinear antiferromagnetic ordering in GdCu$_2$ \cite{GdCu2_1}. For \EC , on the other hand, only magnetization measurements on polycrystalline samples have been reported so far, suggesting a possible antiferromagnetic ordering \cite{EuCd2}. Here, we reveal that \EC\  exhibits rich non-monotonic Hall responses, including THE characterized by multiple peak structures and a hysteresis loop. In particular, one of the THE components is sharply enhanced at the magnetic transition, indicating a Hall contribution derived from possible non-trivial spin configurations formed at the magnetic DBs.


Thin films of \EC\ were epitaxially grown on \AO (0001) substrates by molecular beam epitaxy. The growth temperature was 350$^{\circ}$C. Eu and Cd were supplied by an effusion cell with a Cd-rich flux ratio (Cd/Eu = 15-44) due to the highly volatile nature of Cd as compared to Eu. The film thickness is designed to be 60 nm. As summarized in the Supplemental Material\cite{SM}, structural characterization by x-ray diffraction (XRD) reveals that the $b$-axis of \EC\ is oriented along the surface normal, while the $ac$-plane Cd honeycomb lattice is aligned to that of the \AO\ $c$-plane. Depending on the orientation of the distorted Eu hexagons with respect to the \AO\ substrate, there are three types of in-plane domains as confirmed by reciprocal space map measurements \cite{SM}. Figure 1(c) shows a cross sectional image of the \EC\ film taken along the $a$-axis by high-angle annular dark-field scanning transmission electron microscopy. The periodic arrangement of Eu and Cd atoms characteristic to the \CC -type structure is clearly resolved. 


Figures 1(d)-1(i) summarize magnetization and transport properties of the \EC\ films. The temperature dependence of the magnetization measured with the magnetic field applied out-of-plane ($B_{\mathrm{out}}$) and in-plane ($B_{\mathrm{in}}$, parallel to \AO[$11\bar{2}0$]) exhibits an onset of antiferromagnetic order at \TN\ = 37 K which is consistent with the previous report on polycrystalline bulk samples \cite{EuCd2}. Below \TN , the magnetization $M$ keeps increasing, indicating that the magnetic structure is not a simple colinear antiferromagnetic ordering. Magnetization curves measured as functions of $B_{\mathrm{out}}$ and $B_{\mathrm{in}}$ at 2 K are presented in Fig. 1(f). \EC\ develops ferromagnetic moments at lower temperature as indicated by a small hysteresis loop around zero field as seen in the inset. This ferromagnetism is suppressed above 20 K (see Supplemental Material \cite{SM}). While the magnetization monotonically increases when the field is applied in-plane, a metamagnetic transition is observed around 1.3 T when the field is applied along the out-of-plane hard axis. The metamagnetic transition is observed up to \TN\ irrespective of the low-temperature out-of-plane ferromagnetism, implying the occurrence of a spin reorientation or a spin-flop transition from the antiferromagnetic ground state. 

The temperature dependence of the resistivity presented in Fig. 1(e) exhibits a kink structure at \TN\ reflecting that the metallic conduction is strongly coupled with the localized spins of Eu$^{2+}$. Figures 1(g) and 1(h) show longitudinal resistivity \rhoxx\ and Hall resistivity \rhoyx\ data measured at 2 K in out-of-plane fields. \rhoxx\ exhibits a peak at the field where the metamagnetic transition occurs, and further increasing $B_{\mathrm{out}}$ leads to a negative magnetoresistance until the spins are fully polarized at around 3.8 T. During such a magnetization process, \rhoyx\ exhibits rich peak features with a hysteresis loop in stark contrast to a rather monotonic change of $M$ against $B_{\mathrm{out}}$ in Fig. 1(f). Figure 1(i) displays the non-monotonic Hall signal after subtracting the sum of the ordinary Hall term \rhoyxOHE\ and the $M$-proportional anomalous Hall term \rhoyxAHE . We note that the longitudinal conductivity \sigmaxx\ of \EC\ lies in the so-called intrinsic region of the conventional AHE scaling plot \cite{AHE_scaling}, and \rhoyxAHE\ here is calculated via $r_{\mathrm{s}}\rho_{xx}^{2}M$ with $r_{\mathrm{s}}$ being determined as a fitting parameter. As discussed below, the non-monotonic Hall signal in \EC\ films consists mainly of THE originating from the real-space spin configuration, and hereafter we denote it as \rhoyxTHE .


Figure 2(a) presents a color map of \rhoyxTHE\ as a function of temperature and out-of-plane field (for the Hall resistivity data at different temperatures, see also the Supplemental Material \cite{SM}).  For clarity, we label the phases below (above) the metamagnetic transition around 1.3 T as Phase I (II) and the forced-ferromagnetic phase above the saturation field around 3.8 T as FM, and the magnetic phase boundaries separating different phases are indicated by dotted lines which are drawn based on the results of the magnetization (marked by solid triangles)and magnetoresistance measurements. Importantly, the Hall response of the \EC\ films is characterized by two unique features. Firstly, there are multiple positive and negative peaks appearing on both sides across zero field as indicated by circles with labels P1, P2, and P3 in Figs. 2(b) and 2(c). These peak structures of \rhoyxTHE\ start to develop clearly below \TN , reflecting their magnetic origin. Moreover, among the \rhoyxTHE\ peaks, the P1 peak appears sharply at the metamagnetic transition field (marked with a solid triangle), and this tendency is observed in the entire temperature range below \TN\ .

A Hall resistivity with multiple peaks has been identified as non-monotonic AHE for various Eu-based magnetic semimetals and semiconductors \cite{EuTiO3,EuP3,EuCd2As2,ECS,EuMg2Bi2}. In those systems, the continuous change of the momentum-space Berry curvature is induced mainly by the formation of Weyl points or their shifting with respect to the Fermi energy during the magnetization process. However, this is not the case for \EC . As shown in the Supplemental Material \cite{SM}, \EC\ exhibits a sign change of AHE with developing the out-of-plane ferromagnetism below 20\ K. This observation itself is interesting because it originates from a drastic modulation of the momentum-space band structure taking place in the presence of ferromagnetism in \EC , which may imply the appearance of topologically non-trivial band features near the Fermi level. We note that a similar temperature-dependent sign change of AHE has been reported for the ferromagnetic Weyl metal SrRuO$_3$ \cite{SrRuO3}. Focusing on the non-monotonic Hall term, on the other hand, the overall peak structures of P1, P2, and P3 remain unchanged even across 20\ K \cite{SM}. The absence of sign changes for those peaks indicates that they do not share the same origin as the intrinsic AHE. Therefore, we can reasonably conclude that the non-monotonic Hall effect is originating in the real-space spin Berry phase rather than in momentum space, and hence, it can be ascribed to THE  (see also Supplemental Material \cite{SM} for additional discussions).

Having clarified the origin of the unique Hall responses as topological spin structures, we now focus on the second feature of \rhoyxTHE\ in \EC , i.e., the pronounced hysteresis loop. Generally, the spin chirality of field-induced spin textures changes its sign upon field reversal \cite{MnSi,MnGe,FeGe,GdPdSi,EuAl4}, leading to a field-antisymmetric THE (\rhoyxTHE($B$)=$-$\rhoyxTHE($-B$)). A hysteretic THE has been reported only in special cases such as hysteretic formation of skyrmions in FeGe thin films \cite{FeGefilm}, transitions between skyrmion and anti-skyrmion phases in Mn$_2$RhSn \cite{Mn2RhSn}, and ferromagnet-based systems with interfacial Dzyaloshinskii-Moriya interaction \cite{SROSIO,MagTI,MagTI2,MagTI3}. In \EC , on the other hand, ferromagnetism develops below 20\ K, and the hysteretic behaviour of \rhoyxTHE\ can be straightforwardly associated with the polarized out-of-plane moments. However, we should emphasize that the hysteresis cannot be explained solely by the additional AHE induced by the out-of-plane ferromagnetic moments. As seen in Fig. 2(b), the hysteresis loop is particularly enlarged around the P1 THE peak due to its asymmetric appearance depending on the field scan directions; in the field-decreasing (increasing) sweep at 2 K, the P1 peak is suppressed on the positive (negative) field side, and it is sharply enhanced on the negative (positive) field side, respectively. This highlights the sensitive coupling between the P1 THE peak and the remanent ferromagnetic moment. 

Interestingly, such coupling between THE and ferromagnetism is not observed for the other THE peaks (P2 and P3), which appear on both sides of the field regardless of the scan direction and the metamagnetic transition. These two different THE components can be also distinguished by measuring their field-angle dependence \cite{SM}. We observe that the amplitude of the P1 peak is quickly diminished as the metamagnetic transition is suppressed with tilting the field toward the in-plane direction, while the P2 peak survives up to high tilting angles (see the Supplemental Material \cite{SM} for details). From these findings, the complex Hall responses of \EC\ can be separated into three components as illustrated in Figs. 2(d) and 2(e); (i) the THE component appearing without hysteresis before the full polarization of the moments along the out-of-plane direction has been reached (P2 and P3), (ii) the THE component which is sharply enhanced at the metamgenetic transition with coupling to the ferromagnetism (P1), and (iii) the ferromagnetism-induced additional AHE component appearing below 20 K. As discussed in the following, we propose that the former THE component represents the bulk contribution and the latter component the DB contribution. 

To clarify the DB-derived nature of the P1 THE peak, we have performed minor loop measurements. Figure 3 presents the Hall term \rhoyx$-$\rhoyxOHE\ measured by scanning the field from $-5$ T to a certain maximum value $B_{\mathrm{max}}$ and then back to $-5$ T. $B_{\mathrm{max}}$ is varied from $-2.2$ T to $2.2$ T so that the minor loops cover the peak and hysteretic features of \rhoyx . To evaluate the hysteretic behavior, we have also extracted the loop term \Deltarhoyx\ defined as the difference between the field-increasing sweep and the field decreasing sweep as shown in Figs. 3(d)-3(f). \Deltarhoyx\ corresponds to the summation of the hysteresis contribution from the P1 THE term and the additional AHE term as illustrated in Fig. 2(d). A striking observation is that there is a region of negative \Deltarhoyx\ in the minor loops of $B_{\mathrm{max}}$ = 0.9, 1.3, 1.5, and 1.6 T as shown in the inset in Figs. 3(e) and 3(f). The negative \Deltarhoyx\ indicates an increase of THE during the field-decreasing sweep as compared to the field-increasing sweep. In particular, the amplitude of negative \Deltarhoyx\ is the largest for $B_{\mathrm{max}}$ = 1.3 T where the magnetic field is reversed during the metamagnetic transition. These observations reveal that the promoted formation of DBs by the minor loop scans leads to a larger amplitude of THE, highlighting the THE contribution of the DBs. We also note that the constant discrepancy (indicated by a two-headed arrow) in the peak amplitude between the full loop and the minor loops with $B_{\mathrm{max}}$ = 0, 0.5, 0.9, and 1.3 T corresponds to the \DeltarhoyxAHE\ term. \DeltarhoyxAHE\ is suppressed in the minor loops for $B_{\mathrm{max}} > 1.3$ T as shown in Fig. 3(c), which means that the metamagnetic transition also promotes the reversal of the out-of-plane moments. 

To further demonstrate the DB-driven nature of the P1 THE peak, we have also examined its dependence on different field cooling processes, which can effectively modulate the DB density. Figure 4 presents \rhoyx$-$\rhoyxOHE\ measured after experiencing two different field cooling paths from above \TN\ to 2 K; in Path 1 the out-of-plane field was gradually increased from 0 T at \TN\ to \BFC\ at 2 K, while in Path 2 the constant field \BFC\ was applied from above \TN\ to 2 K as shown in the top panels. After reaching \BFC\ at 2 K, field scans towards $+$5 T and $-$5 T were performed. \rhoyx$-$\rhoyxOHE\ of the \BFC\ = 0 T case presented in Fig. 4 (a) reflects the initial magnetization curve after zero field cooling, and exhibits the P1 peak on both positive and negative field sides, in contrast to the full loop shown in gray for comparison. This can be interpreted to reflect the absence of the ferromagnetic component suppressing the DB-driven THE under zero field cooling. 

For evaluating the modulation of DB-driven THE for each \BFC\, we have specifically taken the change of the Hall resistivity from that of the \BFC\ = 0 T case (\rhoyx-\rhoyx$_{,B_{\mathrm{FC}} = 0 T}$) as presented in the lower panels. When \BFC\ is much lower than the metamagnetic transtion field around 1.3 T such as in the case of \BFC\ = 0.7 T in Fig. 4(b), both of the field cooling along Path 1 and Path 2 result in enhanced amplitude of the P1 peak. When \BFC\ is 1.3 T, on the other hand, Path 1 and Path 2 exhibit a contrasting behaviour. As shown in Fig. 4(c), Path 1 for \BFC\ = 1.3 T following closely along the magnetic phase boundary between I and II leads to significant enhancement of the P1 peak, which even exceeds that of the full loop. In contrast, Path 2 following the outside of the phase boundary results in suppression of the P1 peak. These observations in the \BFC\ = 1.3 T case clearly indicates the strong relevance of THE to the DB density formed during different field cooling paths in the \EC\ films. Further increase of \BFC\ strengthens the development of the out-of-plane moments, suppressing DB-driven THE for both Path 1 and Path 2 as shown in Fig. 4(d) (see also the Supplemental Material \cite{SM} for the results of other \BFC\ cases, and also the field cooling dependence of the magnetization curve).

Finally, we would like to discuss the possible origin of the THE signals in \EC . THE appears as a bulk effect when non-coplanar spin textures with finite scalar spin chirality are stabilized by the external field. If we simply assume a non-colinear ordering such as 120$^{\circ}$-spin ordering for the Eu triangular lattice of \EC , however, the field-induced spin canting generates scalar spin chirality only locally and cancels it out globally due to the contribution of opposite signs from the adjacent triangles. To realize a non-vanishing scalar spin chirality as bulk effect in a triangular lattice system, either strong coupling to the spin-orbit interaction or an incommensurate non-coplanar ordering is required to break the balance of this cancellation \cite{MultiQ}. As proposed in the theory \cite{MultiQ} and experimentally verified by the realization of a skyrmion phase in \AB -type \GPS \cite{GdPdSi}, the presence of further neighbor interactions on a triangular network of Heisenberg-type spins can lead to incommensurate spiral textures with multiple-$Q$ vectors. It is worth noting that the nearest-neighbor Eu sites in \CC -type \EC\ are the interlayer Eu-Eu sites rather than the intralayer triangular sites. The ferromagnetism below 20 K also indicates the presence of ferromagnetic interaction between some Eu-Eu sites. Altogether, it is highly likely that not just a simple two-dimensional antiferromagnetic ordering within the triangular lattice plane but a more complex three-dimensional non-coplanar ordering is realized in \EC , accounting for the observations of the P2 and P3 THE peaks.  

Opposed to the bulk THE, the particularly sharp THE at the metamagnetic transition is attributed to the contribution from the magnetic DBs. In contrast to the bulk scalar spin chirality on the triangular lattice, that of the local DBs is expected to survive owing to the broken symmetry. The fact that the peak amplitude of P1 is suppressed when the out-of-plane moments remain polarized under higher fields, and that it is enhanced when the moments are weakened at the switching field, proves that the spin configuration realized at the DBs is crucial for the appearance of the THE. Such a dependence of THE on ferromagnetism actually resembles the THE observed in ferromagnet-based heterostructures with interfacial DM interaction  \cite{SROSIO,MagTI,MagTI2,MagTI3}. There, a THE peak structure appears only at the coercive field of the ferromagnetic layer, where the DM interaction comes in to realize chiral DBs or skyrmions. Since interfacial or bulk DM interaction is not expected for the present case, we speculate that instead the presence of frustration-induced non-coplanar spin textures in \EC\ plays a vital role in realizing non-trivial DBs hosting finite scalar spin chirality. 

In summary, we have succeeded in the film growth of the \CC -type frustrated magnet \EC\ and have revealed its rich Hall responses, including a temperature-dependent sign change of AHE and multiple THE peaks accompanied by a pronounced hysteresis loop. One of the THE peaks is sharply enhanced at the metamagnetic transition field, indicating the formation of magnetic DBs hosting finite scalar spin chirality. Importantly, compared to the bulk AHE and THE, the DB contribution appears as the leading term. For the formation of non-trivial DBs which give such a dominant Hall response, it is expected that the presence of possible non-coplanar spin textures within the frustrated Eu network is essential. The determination of the actual magnetic structure realized in \EC\ is highly desired to further clarify its unique Hall responses.

We thank M. Kawasaki for fruitful discussions and also the help in parts of the low-temperature magnetotransport measurements. We also thank N. Kanazawa, H. Ishizuka, H. Oike, H. Sakai, T. Nakajima, Y. Yamasaki, K. Matsuura, and F. Kagawa for fruitiful discussions. This work was supported by JST FOREST Program grant no. JPMJFR202N, Japan; and by Grant-in-Aids for Scientific Research JP21H01804, JP22H04471, JP22H04501, JP22K18967, JP22K20353, and JP23K13666 from MEXT, Japan.

\begin{figure*}
\begin{center}
\includegraphics[width=14cm]{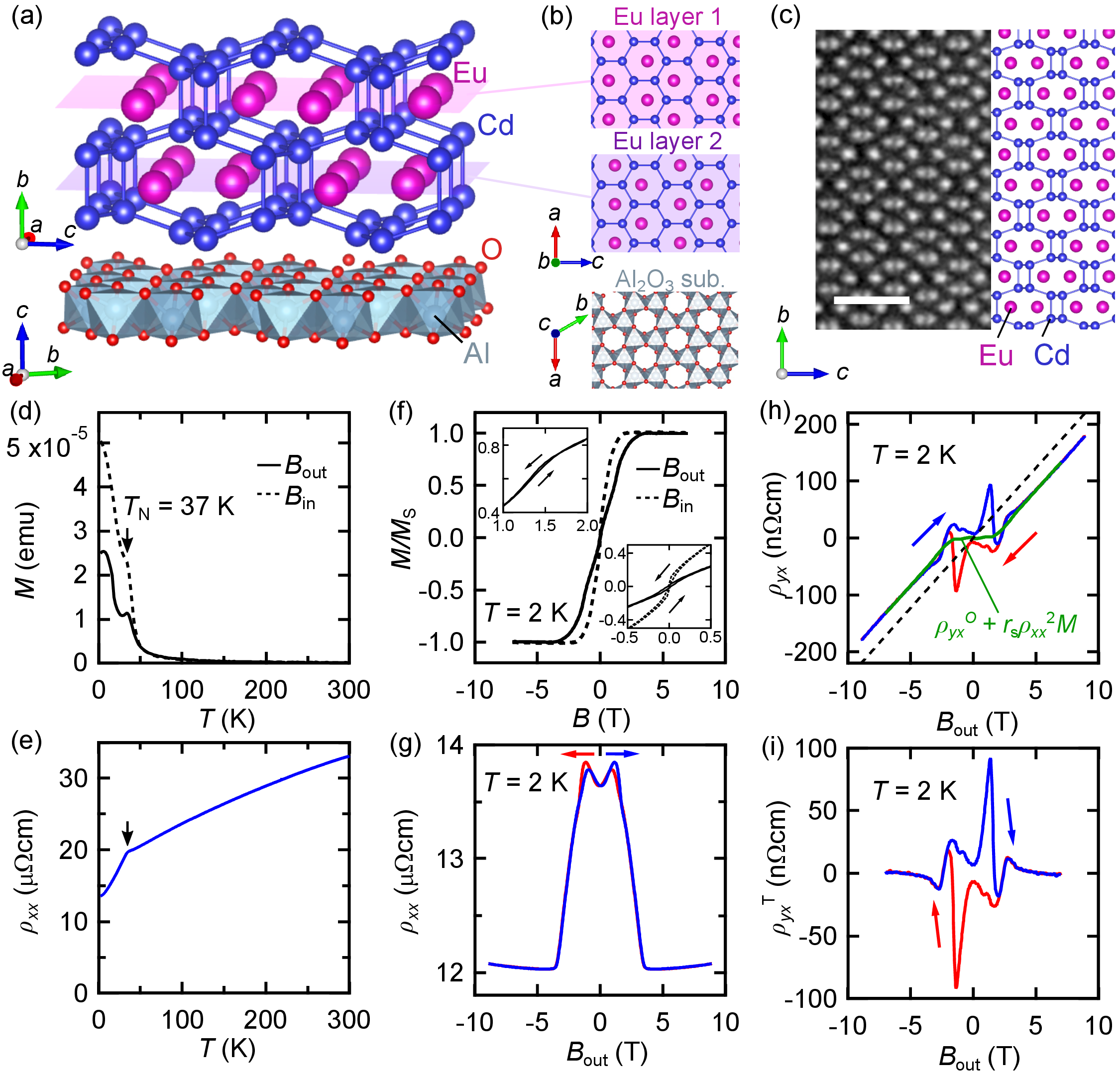}
\caption{
Structural, magnetization, and transport characterization of the \EC\ films. (a) Crystal structure and orientation of \CC -type \EC\ grown on the \AO (0001)  substrate. The Eu distorted triangular lattices and the Cd buckled honeycomb lattices are alternately stacked along the $b$-axis. (b) In-plane epitaxial relation between \EC\ and \AO. Eu atoms are displaced oppositely with respect to the center of the Cd hexagon between adjacent Eu layers, preserving the inversion symmetry. (c) Cross-sectional image of the \EC\ film (left), and atomic arrangement of Eu and Cd atoms in the \CC -type structure viewed along the $a$-axis (right). The scale bar corresponds to 1 nm. (d) Temperature dependence of the magnetization ($M$) measured in a magnetic field of 0.1 T applied in the out-of-plane (\Bout) and in-plane (\Bin) directions. (e) \Bout\ and \Bin\ dependence of the magnetization normalized by the saturation value $M_{\mathrm{s}}$. The insets show magnified views of the hysteresis loop around zero field and the metamagnetic transition at \Bout$ = 1.3$ T. (f) Temperature dependence of the longitudinal resistivity \rhoxx. \Bout\ dependence of (g) \rhoxx\ and (h) Hall resistivity \rhoyx. (i) Topological Hall signal (\rhoyxTHE) extracted by subtracting the ordinary Hall term (\rhoyxOHE) and the $M$-proportional anomalous Hall term (\rhoyxAHE\ = $r_{\mathrm{s}}\rho_{xx}^{2}M$) from the raw data in (h). 
}
\label{fig1}
\newpage
\end{center}
\end{figure*}

\begin{figure*}
\begin{center}
\includegraphics[width=15cm]{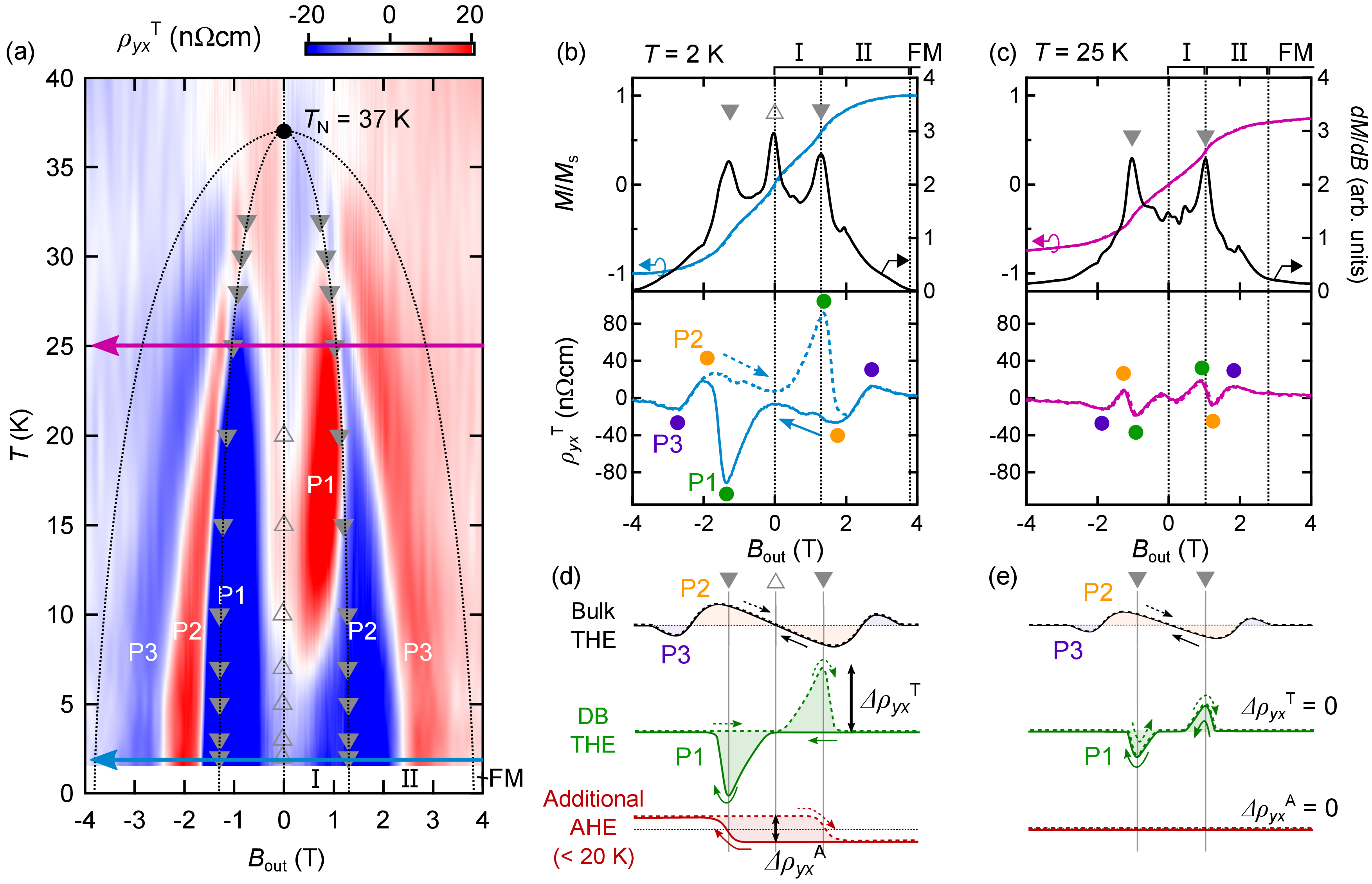}
\caption{
Topological Hall effect enhanced at the metamagnetic transitions. (a) Color map of the topological Hall resistivity \rhoyxTHE\ as a function of temperature $T$ and out-of-plane magnetic field \Bout . Only data of the field-decreasing sweep from 9 T to $-9$ T are shown. Open and solid triangles denote the field position of the magnetic transition, and dotted lines show the magnetic phase boundaries between I, II and FM. Normalized magnetization and \rhoyxTHE\ measured upon sweeping the field at (b) 2 K and (c) 25 K, as indicated by the arrows in (a). The \rhoyxTHE\ peak induced by the metamagnetic transitions is labeled P1 with a green solid circle, while the bulk THE peaks are labeled P2 (yellow circle) and P3 (purple circle). Schematic illustrations of the decomposed Hall contributions from different origins at (d) 2 K and (e) 25 K. The amplitude of the hysteresis loop in the DB-driven THE and the additional AHE below 20 K is denoted by \DeltarhoyxTHE\ and \DeltarhoyxAHE , respectively.
}
\label{fig2}
\end{center}
\end{figure*}

\begin{figure*}
\begin{center}
\includegraphics[width=15cm]{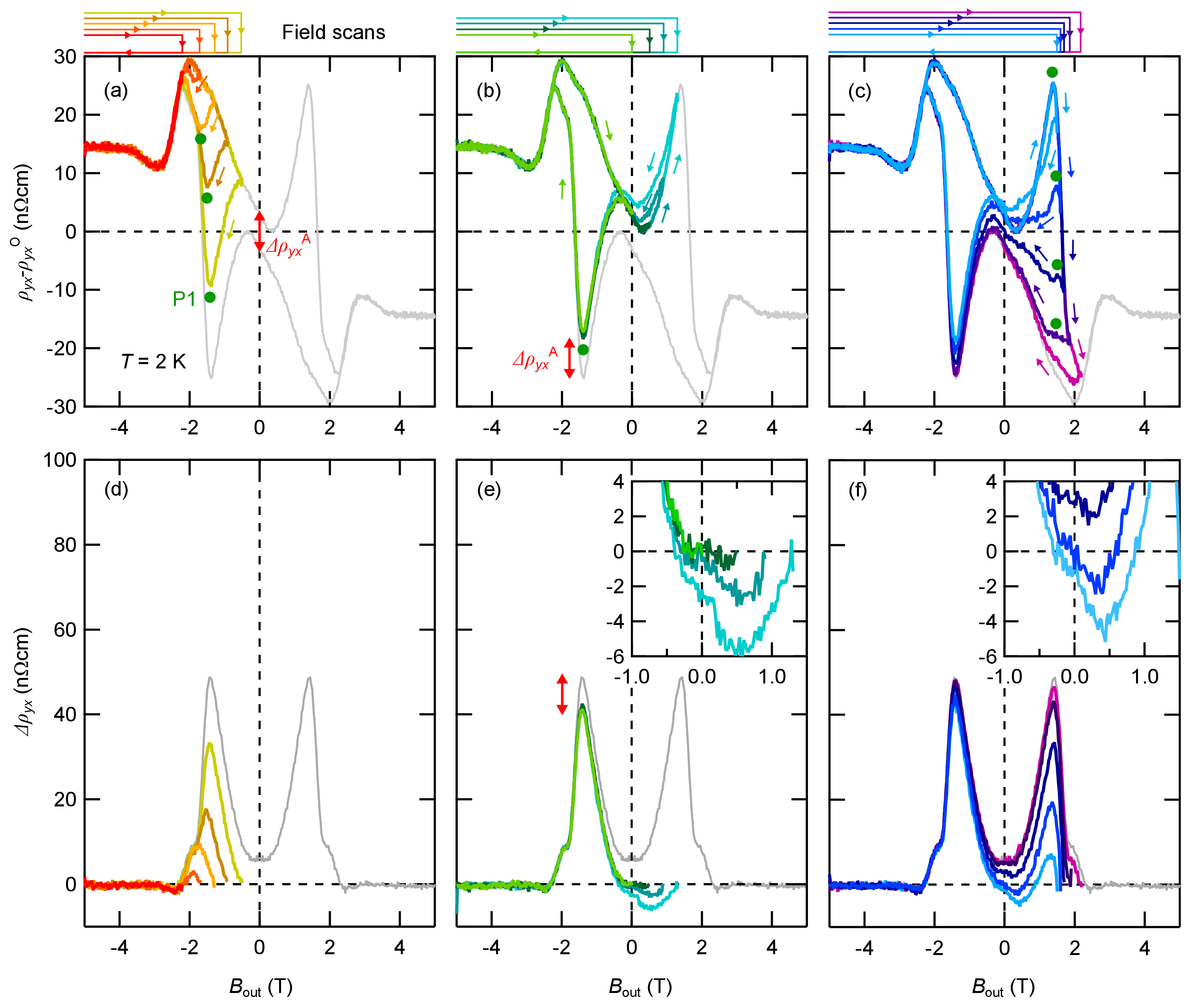}
\caption{
Minor loop measurements of the Hall resistivities after poling at $-5$ T. The Hall term \rhoyx$-$\rhoyxOHE\ measured for the minor loops with $B_{\mathrm{max}}$ = $-2.2, -1.7, -1.3, -0.9$ T, (b) $B_{\mathrm{max}}$ = $0, 0.5, 0.9, 1.3$ T, and (c) $B_{\mathrm{max}}$ = $1.5, 1.6, 1.7, 1.9, 2.2$ T. The full hysteresis loop is shown in gray for comparison. The arrows on top of the panels (a)-(c) show the field scan direction. The color code corresponds to the color of the respective data. (d)-(f) The loop term \Deltarhoyx\ extracted from the respective \rhoyx$-$\rhoyxOHE\ curves in (a)-(c). $\varDelta\rho_{\mathrm{AHE}}$ is the width of the zero field hysteresis reflecting the additional anomalous Hall contribution from the out-of-plane ferromagnetic moments. Insets in (e) and (f) show the field region with a negative \Deltarhoyx\ observed for the loops with $B_{\mathrm{max}}$ around 1.3 T, which indicates that the enhanced formation of DBs yields a larger P1 THE peak.
}
\label{fig3}
\end{center}
\end{figure*}

\begin{figure*}
\begin{center}
\includegraphics[width=15cm]{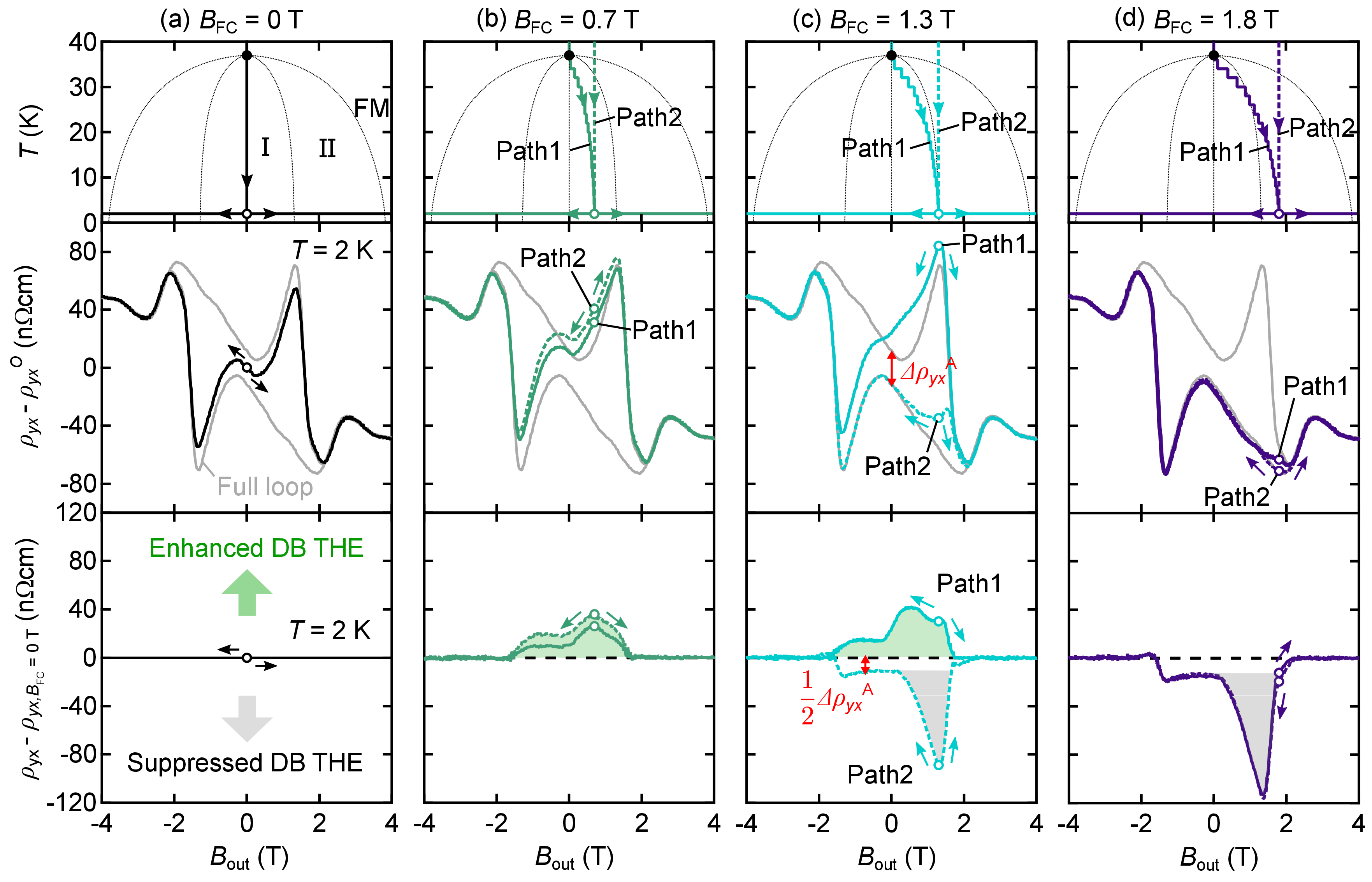}
\caption{
Modulation of THE by different field cooling processes. Top panels show the two different field cooling paths (Path 1 and Path 2) from above \TN\ to 2K and to the field \BFC\ = (a) 0 T, (b) 0.7 T, (c) 1.3 T, and (d) 1.8 T. \BFC\ = 0 T corresponds to the zero field cooling, and Path 1 with \BFC\ = 1.3 T follows closely the magnetic phase boundary between I and II. Hall term \rhoyx$-$\rhoyxOHE\ measured at 2 K after the respective field cooling path is shown in the middle panels. The solid curve presents the result for Path 1 and dashed line for Path 2 for each \BFC . The field scans were performed in both the directions towards $-$5 T and $+$5 T starting from \BFC\ as marked by an open circle. For comparison, the Hall resistivity curve of the full loop scan is also shown in gray. The lower panels show modulation of THE evaluated by subtracting \rhoyx\ of the \BFC\ = 0 T case from that of respective \BFC\ case. The peak amplitude of DB THE is particularly enhanced when the field cooling path follows the inside region of the phase boundary between I and II.      
}
\label{fig4}
\end{center}
\end{figure*}

%
\clearpage
\end{document}